# Bonding, Conductance and Magnetization of Oxygenated Au Nanowires

C. Zhang, R. N. Barnett, U. Landman

School of Physics, Georgia Institute of Technology, Atlanta, GA, 30332-0430

## Abstract

Spin-density-functional calculations of tip-suspended gold chains, with molecular oxygen, or dissociated oxygen atoms, incorporated in them, reveal structural transitions for varying lengths. The nanowires exhibit enhanced strength for both oxygen incorporation modes, and upon stretching tip atoms join the wire. With incorporated molecular oxygen the wire conductance is about $1(2e^2/h)$, transforming to an insulating state beyond a critical length. The nanowire conductance with embedded oxygen atoms is low, $0.2\ (2e^2/h)$, and it develops magnetic moments localized on the oxygens and the neighboring Au atoms.

Metal nanowires (NWs) are among the best studied nanoscale systems, exhibiting emergent structural, mechanical, electronic and chemical properties that differ from those of the corresponding materials in their bulk form [1, 2]. In particular, gold nanowires have been the topic of continued interest since the early description of their formation and properties through molecular dynamics simulations of the elongation of a contact between a Ni tip and an Au surface [3]. Coupled with advances in the field of nanocatalysis [4], focused on the surprising catalytic activity of small gold clusters (in particular, the low temperature oxidation of CO by gold clusters with up to 20-30 atoms), recent investigations explored chemical modifications of gold nanowires [5 – 12]

From experiments on Au nanowires it has been concluded [9] that oxygen atoms (rather than molecular oxygen) incorporate and reinforce the wires thus allowing the drawing of longer gold chains. In experiments at 4.2K, as well as 40K, the conductance histogram of the oxygen-incorporated Au suspended chains exhibits a peak at $1G_0 = 2e^2/h$, as in the case for bare Au nanowires where this peak corresponds to the last conductance plateau before breaking. A small peak at $0.1G_0$ has been also recorded at 40K, but no structural assignment corresponding to this peak was made [9]. The description of the structure underlying the 4.2K and 40K conductance peak at $1G_0$ in terms of incorporated atomic oxygen into the Au chain was made [9] on the basis of a theoretical calculation of the band-structure of a one-dimensional alternating -Au-O-Au-O- infinite chain [7], where it has been concluded that the conductance of such a chain will be $1G_0$. The enhanced strength of a supported gold nanowire (chain) upon the insertion of an oxygen atom has been addressed also by recent theoretical calculations [11,12].

Here we report on first-principles density-functional theory (DFT) calculations of the properties of gold nanowire chains suspended between opposing pyramidal Au tips, with molecular oxygen (mode (i)) or dissociated oxygen atoms (mode (ii)) incorporated in the nanowires. Through determination of the electronic structure, bonding characteristics, and optimal atomic arrangements, as a function of the length of the wires, we find an enhancement of the nanowire strength for both modes of oxygen incorporation. Furthermore, from electronic conductance calculations (using a spin-dependent non-equilibrium Green's function, SD-NEGF, technique [13]) we find that the conductance of mode (i) wires is near $1G_0$ up to a critical length beyond which the wire transform into an insulating state. Consequently, the molecular incorporation mode (i) is assigned to the main conductance peak measured in the aforementioned experiments (at both 4.2K and 40K) [9]. In contrast, the conductance of mode (ii) nanowires (dissociative oxygen incorporation) is found to be low (up to $0.2G_0$) with a weak dependence on the wire length, and it may be associated with the low conductance peak measured only at the higher temperature (40K). We also predict that in the ground state of the atomic oxygen incorporation mode (ii) the symmetry between states with opposite spin-directions is broken, leading to the development of magnetic moments localized on the oxygen atoms and the neighboring Au atoms of the nanowire. These moments originate from hybridization between the gold d states and the oxygen p orbitals, with partial charge transfer to the oxygens. On the other hand, in the molecular incorporation mode (i) charge transfer, $\delta q$, into the molecule's $2\pi^*$ (e.g. $\delta q = 1.14e$ for state A in Figure 1(i)) quenches the paramagnetism of the free $O_2$ molecules, resulting in a non-magnetic nanowire .

To gain insights about the nature of bonding, atomic arrangements, structural and chemical changes, forces, and electronic transport in Au nanowires formed upon separation of the leads with molecular or atomic oxygen incorporated in the wires, we performed density functional theory (DFT) electronic structure calculations [14] and conductance calculations combining DFT and the non-equilibrium Green's function (NEGF) technique [15]. The DFT calculations include the generalized gradient approximation (GGA) [16], using a plane wave basis (kinetic energy cutoff $E_{cut}$ = 68 Ry) and pseudo–scalar relativistic soft pseudopotentials [17].

The electronic and atomic structures of the contact region (which includes the nanowire and two opposing gold tips of pyramidal shape, see configuration (a) in Fig.1(ii)) where determined for different values of $L_0$. In structural optimizations, all the atoms in the contact region, except for those in the outermost layers (one on each of the tips), were allowed to fully relax for each value of $L_0$; all atoms in the leads connecting to the contact region are held fixed in their bulk positions. We found that a change of $L_0$ expresses itself essentially entirely in the gap (nanowire region) only. These relaxed configurations are subsequently used in the NEGF transport calculations that employ semi-infinite fcc (110) leads (consisting of fcc stacked alternating 16- and 9-atom layers) connected through the contact region. In the atomic leads we used an atomic basis with the same orbitals per atom as those used in the construction of the pseudopotential [17] (with the 6p orbitals added for gold), which are then expanded in a plane wave basis with $E_{cut}$ = 68 Ry.

Results for the energy changes ($\Delta E$) upon elongation of the distance between the tips ($\Delta L$) for two modes of oxygen incorporation in the nanowires [(i) molecular ($O_2$), and (ii) dissociative (atomic)], are given in Fig. (1i) and 1(ii), respectively. For the molecular incorporation [18] we start from $L_0 = 17.19$ Å, where the wire consists of the two Au tip apex atoms (with a distance $d_{Au}(1,2) = 5.08$ Å) and a slanted O-O molecule (see configuration (A) in Fig. 1(i)), with $d_{O(1)-O(2)} = 1.36$ Å which is larger by 0.1 Å compared to the internuclear distance in the free molecule (1.25 Å), indicating an activated molecular state; we note that $d_{O(1)-O(2)}$ varies by at most ± 0.02 Å between configuration A and E shown in Fig. 1. Formation of this activated state involves charge transfer $\delta q$ into the molecule's $2\pi^*$ (from orbital population analysis we find $\delta q = 1.14e$ for state A in Figure 1(i)), as is the case for gold cluster catalysts (see chapter 1 in ref. [4]). The variation of the total energy as the leads are separated is displayed in Fig. 1(i) together with selected configurations along the stretching path. Similar data corresponding to the dissociative case are shown in Fig. 1(ii), starting from the configuration marked (a) containing 5 Au atoms (including the two tip-apex atoms) and two oxygens ($L_0 = 20.19$ Å, the length of the wire is 11.7 Å, the distance between neighboring Au atoms 2.61 Å, and the Au-O distance is 2.01 Å).

The total energy change is characterized by a step-wise variation, reflected by the spikes in the force curve vs. distance ($F \equiv -\delta(\Delta E)/\delta(\Delta L)$, where $\delta$ denotes a finite difference derivative), shown for the molecular case in the bottom panel of Fig.1. These steps are found to be associated with structural changes, with each step corresponding to the extraction of an additional Au atom from one of the tips, and its incorporation into the nanowire; the numbers in the force curve (e.g. "2 to 3") signify a stretching-induced addition of a gold atom to the wire. This elongation (or "wire drawing") mode is a consequence of the stronger Au-O bond compared to the Au-Au bond, resulting in the observed ability to generate longer wires in the presence of oxygen then in an oxygen free environment [9].

The electronic conductance as a function of elongation of the nanowire is displayed in Fig. 2a. The conductance of the molecular oxygen incorporation mode (i) remains high (close to $1G_0 = 2e^2/h$) for most of the wire drawing process [19], undergoing a sharp transition to a non-conducting state after stage D (see Fig.1(i) where the distance between the two nanowire gold atoms at the tips apexes is 11.68 Å, the average distance between Au atoms is 2.6 Å, $d_{O(1)-O(2)} = 1.34$ Å, $d_{Au-O} = 2.06$ Å, and 2.27 Å). The transition to an insulating wire is illustrated in Fig. 2b, where the isosurface of the density of electrons with energies near the Fermi level, $n(\mathbf{r}; E_F)$, for a conductive state (C) is seen to be extended over the two tips and the nanowire (including the embedded oxygen molecule), while for an insulating state (i.e. for $\Delta L > 5$ Å) the two tips and the nanowire appear to be electronically disconnected, with $n(\mathbf{r}; E_F)$ localizing on the tips and only on part of the nanowire. For the dissociated case (mode (ii) the conductance of the wire is low from the beginning, and it decreases monotonically (see insert in Fig. 2a). Based on these results we assign the molecular oxygen incorporation mode as that corresponding to the main peak (at $1G_0$) in the conductance histogram measured (at both 4.2K and 40K) in ref. 9, rather than the dissociative mode inferred there from earlier electronic structure calculations [7] for a periodic infinite one-dimensional -Au-O-Au-O- chain. The latter (i.e, dissociative mode) may be associated, instead, with the aforementioned small conductance peak measured [9] only at 40K .

To elucidate the effect of the oxygen atom on the transport properties, we examine the density of states (DOS) projected either on the oxygen atoms or on selected gold atoms of the nanowire. In figs. 3a and 3b we show the projected DOS (PDOS) on the oxygen atom closer to the left lead for the molecular (i) and dissociative (ii) oxygen incorporation modes (corresponding to the points marked A and a in Fig.1(i) and 1(ii), respectively). Similarly, we show the PDOS on the middle Au atom for the dissociative mode (ii) in Fig. 3b. For mode (i) the PDOS on the O atom are the same for the two spin directions, with a large contribution from p states, and a smaller one from Au d states, at the Fermi-level (Fig. 3a). On the other hand, for mode (ii) the symmetry of the PDOS for the two spin directions is broken, and the PDOS at $E_F$ is significantly reduced, resulting in low conductance. The PDOS on the central Au atom for mode (i) is also found to be symmetric (not shown in Fig. 3), with the main contribution at $E_F$ coming from d states (and a smaller one of s character). The PDOS for mode (ii) is, again, not symmetric for the two spin directions, resulting in a lowering of the PDOS at $E_F$; here the main contribution is of d-character, with some s, and a small p, contribution (Fig. 3c).

From these results we conclude that: (a) bonding of oxygen to the gold nanowire has some covalent character, involving hybridization of the O(2p) with Au(5d) and 6(s) orbitals, resulting, as aforementioned, in enhanced strength of the "oxygenated" nanowire; (b) splitting of the PDOS between the two spin directions, causes a reduced population at $E_F$ for the dissociative mode ( ii), correlating with the much lower conductance of these nanowires; (c) the broken spin direction symmetry is predicted to result in magnetization of nanowires with oxygen incorporated dissociatively. Indeed, our calculations show formation of significant magnetic moments in the central part of mode (ii) nanowires, with strong d character on the central Au atom and p character on the oxygens (see Fig.4);

the magnetization exhibits some sensitivity to the stretching (stressing) of the nanowire (see insert).

In summary, with first-principles DFT calculations we have shown that incorporation of molecular oxygen in tip-suspended Au nanowire chains enhances the nanowire's cohesive strength via bonding involving charge transfer to the 2π* of the embedded molecule. Structural transitions predicted to occur upon stretching are accompanied by drawing of tip atoms and their incorporation in the nanowire. The conductance of the non-magnetic undissociated molecular oxygen-containing wire is calculated to be close to $1G_0$, in agreement with experiments (at both 4.2K and 40K) [9], with a predicted transition to an insulating state beyond a wire length of about 12Å (see Fig. 2). Incorporation of atomic oxygen is found to result in a low conductance ($0.2G_0$) magnetic nanowire, with moments localized on the oxygens and neighboring Au nanowire atoms. The dissociative mode of the oxygenated Au wire can be associated with the small low conductance peak found experimentally [9] only at 40K. Formation and exploration of the properties of magnetic atomic-oxygen-containing Au nanowires present future experimental challenges.

**Acknowledgement:** This work was supported by the US Department of Energy and the AFOSR. Computations were performed at NERSC, Berkeley, CA, and the Center for Computational Materials Science at Georgia Tech.

[15] In the NEGF method [J. Taylor *et al.,* Phys. Rev. B **63**, 245407 (2001); M. Bandbyge *et al.,* Phys. Rev. B **65**, 165401 (2002)], we take the Hamiltonian in the two semi-infinite leads as bulk-like. Then, the self energies, $\sum_L$ and $\sum_R$, which describe the interactions between the left (L) and right (R) leads with the contact region and the Green's functions (retarded and advanced, $G^{r,a}$) can be calculated. The transmission in the Landauer expression for the conductance can be evaluated as

$T(\varepsilon) = Tr\left[\Gamma_L(\varepsilon)G^r(\varepsilon)\Gamma_R(\varepsilon)G^a(\varepsilon)\right]$, where $\Gamma_{L,R} = \Sigma^r_{L,R} - \Sigma^a_{L,R}$. The generalization of this formalism to include spin, involves the use of spin-DFT, and the above relations hold for both spin directions. For further details see Ref. 13.

[16] J. P. Perdew, K. Burke, and M. Ernzerhof, Phys. Rev. Lett. **77**, 3865 (1996).

[17] N. Troullier and J.L. Martins, Phys. Rev. B **43**, 1993 (1991). In the pseudo-potentials we include the 2s and 2p orbitals for the oxygens and the 5d and 6s orbitals for the Au.

[18] Choice of the initial configuration was motivated by preliminary explorations that led us to conclude that molecular oxygen incorporation between the two outermost tip atoms (structure A in Fig.1(i)) is the most likely one to correspond to the measured wire-drawing and transport measurements. This configuration is supported by a recent study of the incorporation of an oxygen molecule into a silver contact, see Y. Qi *et al.*, Phys. Rev. Lett. **97**, 256191 (2006).

[19] Interestingly, a 5-atom bare Au nanowire, set up with bulk interatomic distances and then fully relax, was found to have a conductance very close to $1G_0$. In general, the high conductance of states A to D in Fig. 2(i) is similar to that of the corresponding suspended bare gold wire, in the optimal geometry before the introduction of oxygen.

# Figure Captions

Fig.1 Total energy variation versus elongation distance of suspended Au nanowires for the molecular((i) and dissociative (ii) oxygen incorporation modes. Selected relaxed atomic configurations are displayed, with the smaller spheres (blue, on line) denoting the oxygen atoms. The elongation force for mode (i) is shown in the bottom panel, with each of the spikes corresponding to a tip atom joining the nanowire. The definitions of $L_{0\,(\text{the}}$ length of the contact region connecting the two leads) and $L_{gap}$ (the nanowire length) are included in configuration (a) of Fig. (ii).

Fig. 2. (a) Variation of the conductance (in units of $G_0 = 2e^2/h$) of a suspended Au nanowire with incorporated molecular oxygen, plotted vs. the wire elongation. The letters on the curve correspond to the configurations shown in Fig.1. The results for mode (ii) nanowires are shown in the inset. (b) Isosurfaces (green, on line) of the density of electrons with energy near $E_F$, $n(\mathbf{r}; E_F)$, for states C (left) and E (right) of mode (i) oxygenated Au nanowires (see also Fig. 1(i)). In state C (E) $L_{gap}$ = 10.01 Å (13.71 Å), the average distance between Au atoms is 2.59 Å (2.72 Å), $d_{O(1)-O(2)}$ = 1.36 Å (1.38 Å), $d_{Au-O}$ = 2.10 Å (2.24 Å) and 2.11 Å (2.05 Å). In the above, $n(\mathbf{r}; E_F) = \sum_{j,s} [1 - 2 | f(j,s) - 0.5 |] |\Phi_{j,s}(\mathbf{r})|^2$, where $\Phi_{j,s}(\mathbf{r})$ and $f(j,s)$ are, respectively, the Kohn-Sham orbital and the Fermi-Dirac distribution function for energy level j with spin $s = \pm \frac{1}{2}$, and a Fermi temperature $k_B T_F = 0.014$ eV. Larger spheres (yellow, on line) depict gold atoms, and smaller ones (red, on line) correspond to oxygen atoms.

Fig. 3. Projected densities of states (PDOS, in arbitrary units) on one of the oxygen atoms for (a) mode (i) (corresponding to configuration D in Fig.1) and (b) mode (ii) (corresponding to configuration a in the inset in Fig.1) of the oxygenated Au nanowires. The PDOS projected on the central Au atom for the mode (ii) nanowire is shown in (c). The Fermi level is at zero. Solid lines denotes p states (blue on line), dotted lines denote d states (red on line), and dashed line denote s states (black on line).

Fig. 4. Magnetic moments per atom (in units of the Bohr magneton, filled squares) for the different atomic regions (numbered as shown at the upper left corner) for configuration (a) (see Fig. 1(ii)) of an Au nanowire with incorporated atomic oxygen (mode (ii)). Insert: total magnetic moments as a function of elongation distance, starting at $\Delta L = 0$ from configuration (a), see Fig. 1(ii). Up triangles, down triangles, and circles, denote the d, p, and s contributions, respectively.

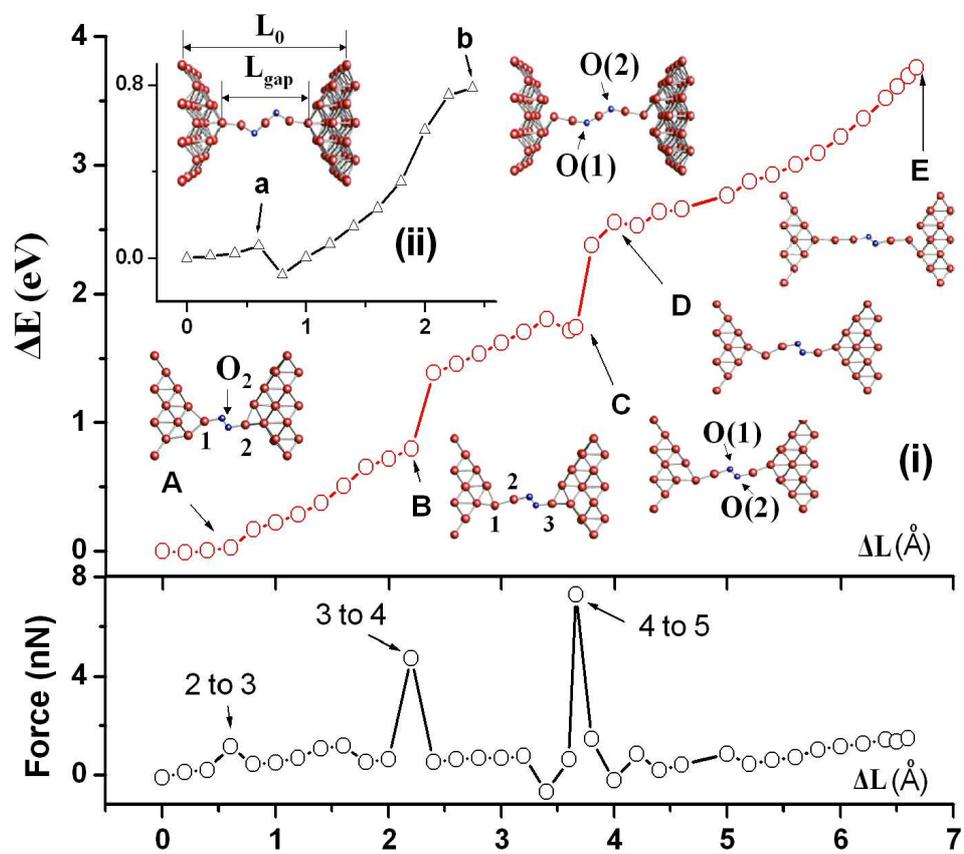

Fig.1

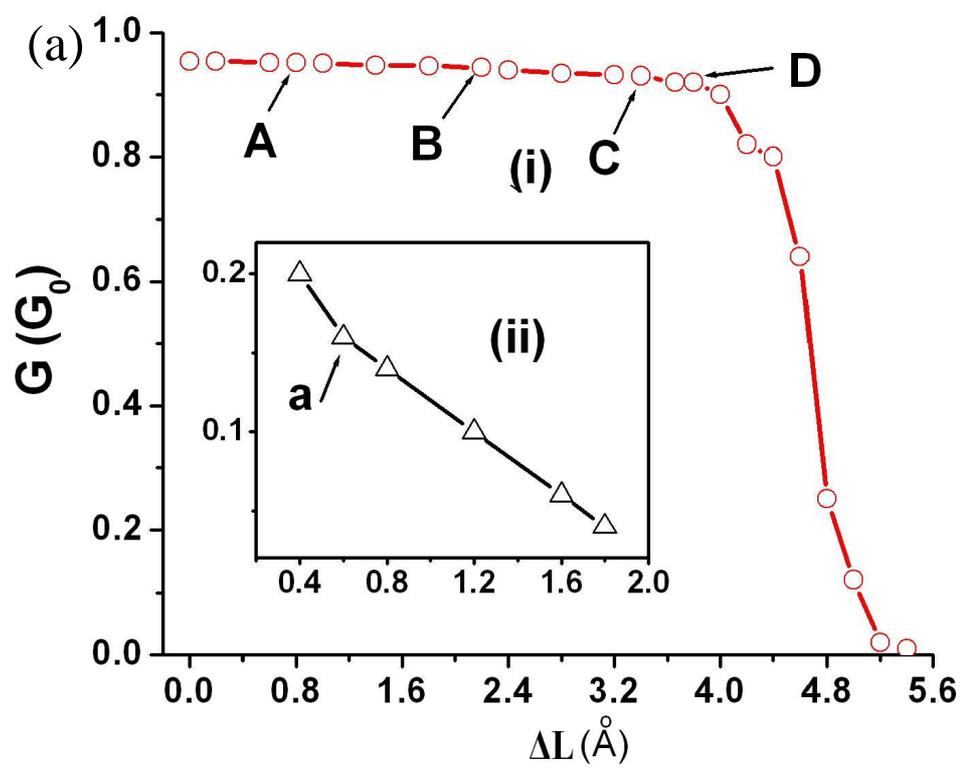

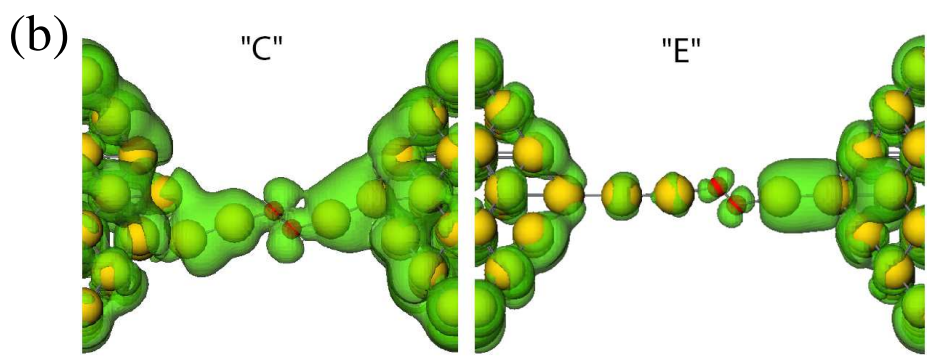

Fig. 2

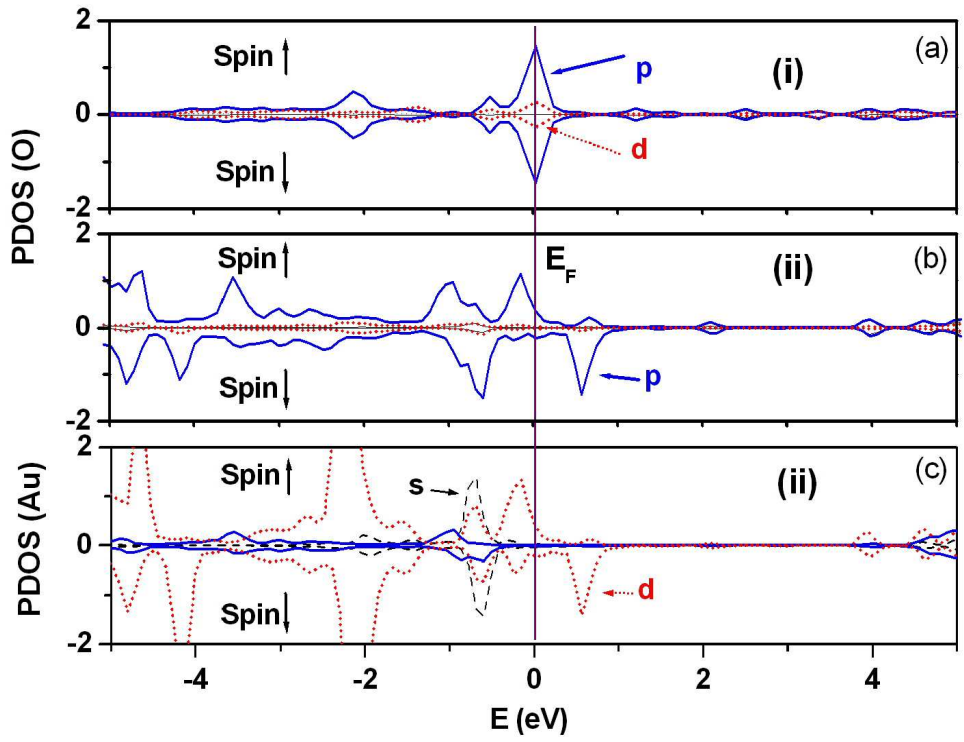

Fig. 3

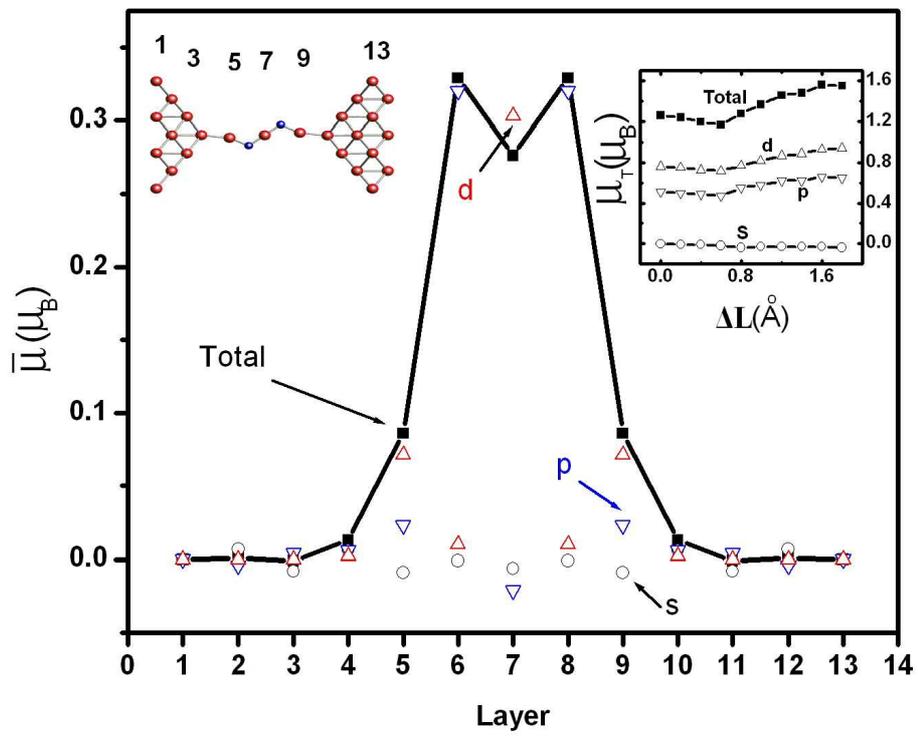

Fig. 4